\documentclass[aps,prl,reprint,superscriptaddress,showpacs,amsmath,amssymb,longbibliography,lengthcheck]{revtex4-1}
\usepackage{graphicx}
\usepackage{mathrsfs}
\usepackage{txfonts}

\begin{document}

\title{A Universal Damping Mechanism of Quantum Vibrations in  
Deep Sub-Barrier Fusion Reactions}

\author{Takatoshi Ichikawa}%
\affiliation{Yukawa Institute for Theoretical Physics, Kyoto University,
Kyoto 606-8502, Japan}
\author{Kenichi Matsuyanagi}
\affiliation{Yukawa Institute for Theoretical Physics, Kyoto University,
Kyoto 606-8502, Japan}
\affiliation{RIKEN Nishina Center, Wako 351-0198, Japan}
\date{\today}

\begin{abstract}
 We demonstrate the damping of quantum octupole vibrations near the
 touching point when two colliding nuclei approach each other in the
 mass-asymmetric $^{208}$Pb + $^{16}$O system, for which the strong
 fusion hindrance was clearly observed. We, for the first time, apply
 the random-phase approximation method to the heavy-mass asymmetric
 di-nuclear system to calculate the transition strength $B$(E3) as a
 function of the center-of-mass distance.  The obtained $B$(E3)
 strengths are substantially damped near the touching point, because the
 single-particle wave functions of the two nuclei strongly mix with each
 other and a neck is formed.  The energy-weighted sums of $B$(E3) are
 also strongly correlated with the damping factor which is
 phenomenologically introduced in the standard coupled-channel
 calculations to reproduce the fusion hindrance.  This strongly
 indicates that the damping of the quantum vibrations universally occurs
 in the deep sub-barrier fusion reactions.
\end{abstract}

% insert suggested PACS numbers in braces on next line\

\maketitle

Heavy-ion fusion reactions are an excellent probe to investigate the
fundamental features of the dynamics for many-body quantum systems.
When a projectile approaches a target, the Coulomb barrier is formed,
because of the strong cancellation between the Coulomb repulsion and the
nuclear attractive force.  Nuclear fusion takes place when the
projectile penetrates through this Coulomb barrier.  At incident
energies in the vicinity of the Coulomb barrier height, called the
sub-barrier fusion, the strong enhancements of fusion cross sections,
compared to the estimations of a simple one-dimensional potential model,
have been observed in many systems.  These enhancements are well
accounted for in terms of the couplings between the relative motion of
the colliding nuclei and the intrinsic degrees of freedom such as
collective vibrations of the target and the projectile~\cite{DHRS98}.
The coupled channel (CC) model, which takes into account this mechanism,
has been successful in describing such
enhancements~\cite{Bal98,Hagino12}.

Recent experiments at extremely low incident energies, called the deep
sub-barrier energies, revealed, however, that steep falloffs of the
fusion cross sections, compared to the estimations of the standard CC
model, emerge in a wide range of mass systems~\cite{Jiang02,das07} (see
Ref.~\cite{Back14} for details). These steep falloff phenomena are often
called the fusion hindrance.  An important quantity for understanding
this fusion hindrance is the potential energy at the touching point of
the colliding nuclei, which is strongly correlated with the threshold
incident energy for the emergence of the fusion hindrance. That is, the
fusion hindrance would be associated with the dynamics in the overlap
region of the two colliding nuclei (see Fig.~1 in Ref.~\cite{ich07-2}).

A theoretical challenge is how to extend the standard CC model to
describe these fusion hindrance phenomena in the overlap region. Two
different models based on assumptions opposite to each other have been
proposed~\cite{Back14}. One is the sudden approach proposed by
Mi\c{s}icu and Esbensen~\cite{mis06,Esb07}. They constructed a heavy
ion-ion potential with a shallow potential pocket considering the Pauli
principle effect acting when two colliding nuclei overlap with each
other.  The other is the adiabatic approach proposed by Ichikawa {\it et
al.}~\cite{ich07-1}. In this approach, neck formations between the
colliding nuclei are taken into account in the overlap region. Based on
this picture, the sudden and adiabatic processes were smoothly jointed
by phenomenologically introducing the damping factor in the coupling
form factor \cite{ich09}.  Later, we showed that the physical origin of
the damping factor is the damping of quantum vibrations of the target
and the projectile near the touching point using the random-phase
approximation (RPA) method for the light mass-symmetric $^{16}$O +
$^{16}$O and $^{40}$Ca + $^{40}$Ca systems~\cite{ich13}.

In this Letter, we show that the damping of the quantum vibrations near
the touching point is a universal mechanism in the deep sub-barrier
fusions and is responsible for the fusion hindrance.  A typical example
optimally suited for this purpose is the recent precise data for the
$^{208}$Pb + $^{16}$O fusion~\cite{das07}. The performances of both the
sudden and adiabatic models have been well tested in this
system~\cite{Esb07,ich09}.  The adiabatic model can reproduce well the
experimental data rather than the sudden model for the fusion
hindrance. To discriminate which model is a better description, we here
show the physical origin of the damping factor introduced in
Ref.~\cite{ich09} in the heavy-mass asymmetric $^{208}$Pb + $^{16}$O
system.

In the standard CC model (and the sudden model), the vibrational modes
of the individual colliding nuclei are assumed not to change, even when
the two nuclei strongly overlap with each other. However, as shown in
Ref.~\cite{ich13}, the single-particle wave functions are drastically
changed by level repulsions, which are associated with the neck
formations.  We apply the RPA method to the heavy-mass asymmetric
system, $^{208}$Pb + $^{16}$O, and show that these mechanisms lead to
damping of quantum vibrations in the colliding nuclei near the touching
point.  This is exhibited by a drastic decreases of the $B($E3)
strengths carried by low-energy RPA excitation modes.

To illustrate our main idea, we first discuss the Nilsson diagram for
protons as a function of the center-of-mass distance, $R$, in the
$^{208}$Pb + $^{16}$O system.  We calculate the mean-field potential for
the $^{208}$Pb + $^{16}$O system using the folding procedure with the
single Yukawa function~\cite{Bol72}.  Before the touching point, we
assume the spherical shape for both nuclei.  After the touching point,
we describe the nuclear shapes with the reflection-asymmetric
lemniscatoids parametrization \cite{Roy85}. 
(The parametrization dependence is negligible, because in this Letter 
we do not discuss the strongly overlapping region.)  
Based on these densities, we also calculate the Coulomb potential.  We
use the radius for the proton and neutron potentials, $R_0$, with $R_0 =
1.27A^{1/3}$ fm, where $A$ is the total nucleon number.  The depths of
the neutron and proton potentials for individual $^{16}$O and $^{208}$Pb
nuclei, $V_T$ and $V_P$, are taken from Ref.~\cite{Mol95}.  In the
folding procedure, we smoothly joint the two different depth parameters
of the mean-field potentials for $^{18}$O and $^{208}$Pb by the function
$V_0(z)=\frac{1}{2}\left[(V_T-V_P)\cdot
\mathrm{erf}\{(z-z_c)/\mu\}+(V_T+V_P)\right]$, where $z_c$ denotes the
center position between the two surfaces of the colliding nuclei and
$\mu$ denotes the smoothing parameter. We take $\mu=0.8$ fm, which is
the same as the diffuseness parameter of the single-particle potential.
In the calculations, the origin is located at the center-of-mass
position of the two nuclei.

Using the obtained mean-field potentials, we solve the axially-symmetric
Schr\"odinger equation with the spin-orbit force. The details of the
model and the parameters are similar to Refs.~\cite{Bol72,Mol95}.  Then,
the $z$ component of the total angular momentum, $\Omega$, is the good
quantum number.  Note that the parity is not a good quantum number
because the mean-field potential for the whole system breaks the
space-reflection symmetry.  We expand the single-particle wave functions
in terms of the deformed harmonic-oscillator bases in the cylindrical
coordinate representation.  The deformation parameter of the basis
functions is determined so as to cover the target and the
projectile. The basis functions with energies lower than 26
$\hbar\omega$ are taken into account.

\begin{figure}[htbp]
 \includegraphics[keepaspectratio,width=\linewidth]{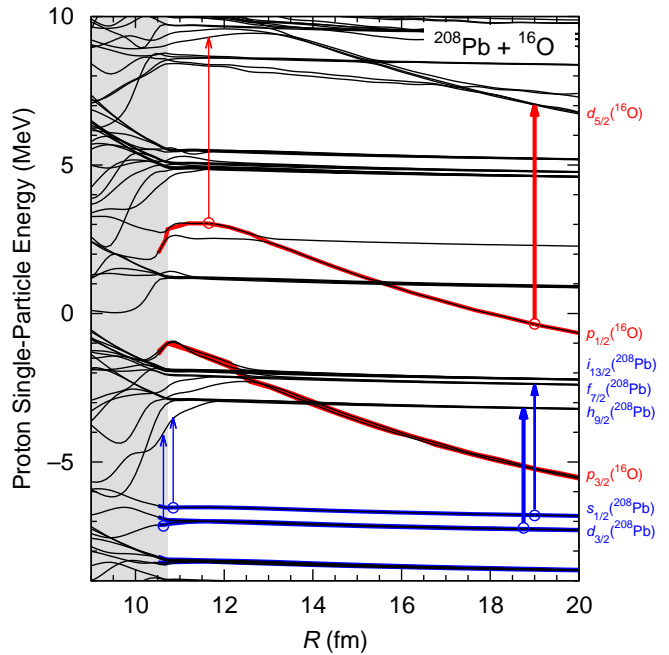}
 \caption{(Color online) Nilsson diagram for the $^{208}$Pb + $^{16}$O
 syatem as a function of $R$. The light gray (red) thick lines represent
 the occupied $p_{1/2}$ and $p_{3/2}$ states in $^{16}$O. The dark gray
 (blue) thick solid lines show the occupied states in $^{208}$Pb. The
 gray area indicates the overlap region of the colliding nuclei. The
 arrows represent the main particle-hole excitations constituting the
 RPA modes.}  \label{nil}
\end{figure}

Figure~\ref{nil} shows the Nilsson diagram as a function of the
center-of-mass distance $R$.  In the figure, we can see extremely strong
Coulomb effect of $^{208}$Pb on $^{16}$O. The single-particle $p_{1/2}$
and $p_{3/2}$ states in $^{16}$O are shown by the (red) thick solid
lines. Even at the large separation distance $R=20$ fm, the energies of
these two states are higher than the Fermi energy of the $s_{1/2}$ state
in $^{208}$Pb. The miss-match of the two Fermi levels between $^{16}$O
and $^{208}$Pb occurs due to the strong Coulomb effect. At an infinite
separation distance, the energies of the $p_{1/2}$ and $p_{3/2}$ states
for $^{16}$O are $-5.88$ and $-10.7$ MeV, respectively. Thus, at $R=20$
fm, the depth of the mean-field potential for $^{16}$O becomes shallow
by about 5 MeV due to the Coulomb effect from $^{208}$Pb.

The single-particle energies of the $p_{1/2}$ and $p_{3/2}$ states in
$^{16}$O remarkably increase with decreasing $R$ due to the increasing
Coulomb effect from $^{208}$Pb.  Then, many level crossings and
repulsions between the energy levels of $^{16}$O and $^{208}$Pb occur.
With decreasing $R$, the energy of the $p_{1/2}$ state becomes positive
around $R=18$ fm, that is, it changes into a {\it resonance} state, but
there is still a sufficiently high Coulomb barrier.  After that, it
goes across the $f_{5/2}$ and $p_{3/2}$ states of $^{208}$Pb around
$R=16$ fm and the $p_{1/2}$ state of $^{208}$Pb around $R= 13$ fm.
Below $R=13$ fm, the Coulomb barrier becomes lower due to the
attractive nuclear mean-field potential.  Then, the strong mixture of
the single-particle states between $^{16}$O and $^{208}$Pb starts in
many levels, which causes many level splittings seen in the Nilsson
diagram.
  
We now solve the RPA equation at each $R$ for the mass-asymmetric
$^{208}$Pb + $^{16}$O system. We calculate the first excited $3^{-}$
(octupole vibrational) states of $^{16}$O and $^{208}$Pb, which give the
main contributions in the standard CC calculations.  We can easily apply
the RPA method to the di-nuclear system, because its the wave function
is described with a one-center Slater determinant.  We take the
single-particle levels for each neutron and proton up to 200th and the
coherent superposition of all one-particle one-hole states with
excitation energies below 30 MeV. We follow the diabatic single-particle
configuration corresponding to the ground state of $^{16}$O.  The
occupied $p_{1/2}$ and $p_{3/2}$ states in $^{16}$O are represented by
the light gray (red) thick curves in Fig.~\ref{nil}. We carry out the
RPA calculation avoiding immediate vicinities of the level-crossing
points.  We use the density-dependent residual interaction taken from
Ref.~\cite{Shl75} and tune it so that the energy of the spurious
center-of-mass motion becomes zero.  We calculate $B$(E3) values for the
RPA solutions with $\Omega=0$ in individual nuclei using the shifted
octupole operator, $\widehat{Q}_{30}(R-R'_0)$, where $R'_0$ is the
center-of-mass position of the projectile or target nucleus.

\begin{figure}[htbp]
\includegraphics[keepaspectratio,width=\linewidth]{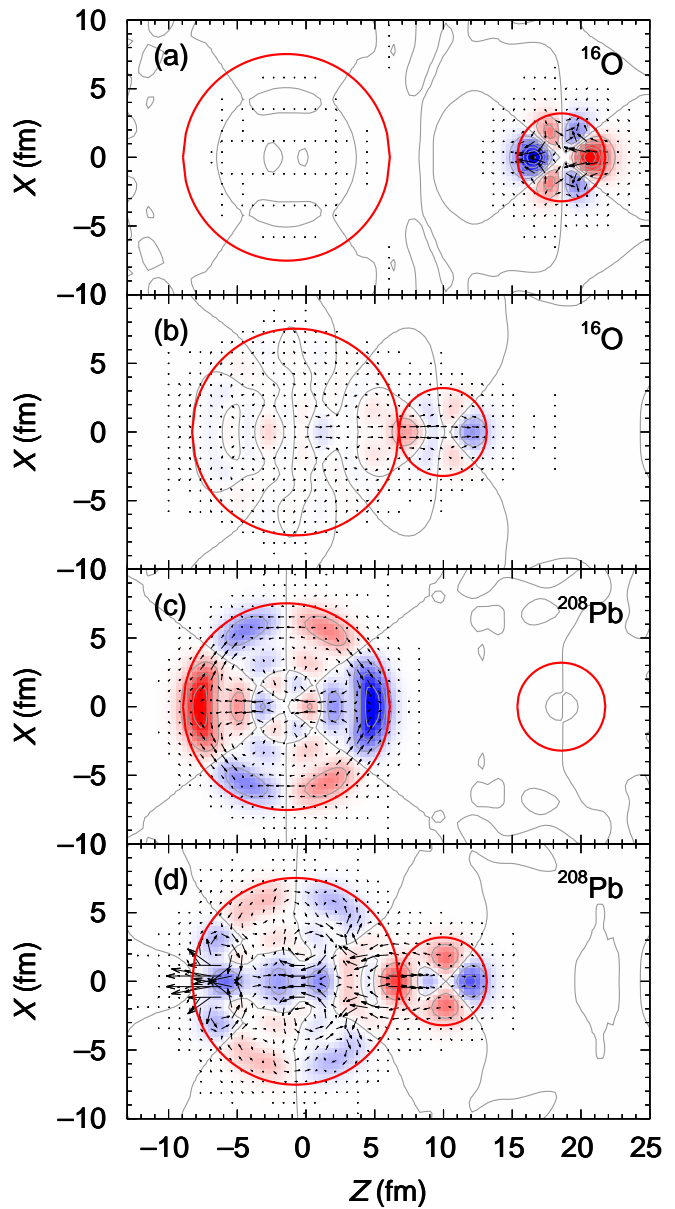}
 \caption{(Color online) Contour maps of the proton transition densities
 and current distributions for the first excited $3^-$ states of
 $^{16}$O and $^{208}$Pb at $R=20$ and 10.73 fm. The contour lines
 correspond to multiples of 0.01 fm$^{-2}$ and 0.005 fm$^{-2}$ for
 $^{16}$O and $^{208}$Pb, respectively.  The arrows represent the
 current density.  The currents and colors are normalized at $R=20$ fm
 in individual nuclei. The (red) thick solid circles indicate the half
 depth of the mean-field potential.}  \label{den}
\end{figure}

At the large separation distance $R=20$ fm, we obtain the first $3^-$
excited states of individual nuclei. The obtained energies and $B$(E3,
$3_1^- \rightarrow 0_1^+$) values are 2.86 MeV and $7.13\times 10^4$
$e^2\cdot {\rm fm}^6$ for $^{208}$Pb and 4.64 MeV and 124 $e^2\cdot{\rm
fm}^6$ for $^{16}$O.  The obtained transition densities and currents for
the first $3^-$ states of $^{16}$O and $^{208}$Pb are depicted in
Figs.~\ref{den} (a) and (c).  At $R=20$ fm, these modes are isolated.
When the two nuclei approach each other, however, these modes start to
fragment into several states.  To evaluate the octupole collective
strengths carried by low-energy excitations, we then calculate the
energy-weighted sum of $B$(E3) strengths.  By checking the spectrum of
all obtained RPA modes as a function of $R$, we determined to take the
sum for octupole excitations with $E \le 4$ MeV and $E \le 6$ MeV for
$^{208}$Pb and $^{16}$O, respectively.

\begin{figure}[htbp]
 \includegraphics[keepaspectratio,width=\linewidth]{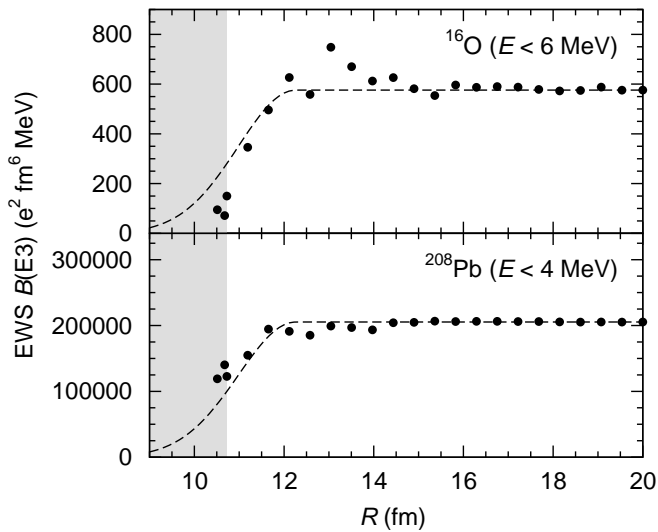}
 \caption{Energy-weighted sums of $B$(E3) for (a) $^{16}$O and (b)
 $^{208}$Pb as functions of $R$. The solid circles show the results of
 the RPA calculations. The dashed curves represent the damping factor
 which well reproduces the experimental data of the fusion cross section
 for $^{208}$Pb + $^{16}$O taken from Ref.~\cite{ich09}. The gray area
 indicates the overlap region of the colliding nuclei.}  \label{damping}
\end{figure}

Figure \ref{damping} shows the $B$(E3) strengths for (a) $^{16}$O and
(b) $^{208}$Pb as functions of $R$.  The calculated values (the solid
circles) drastically decrease near the touching point (the boundary
between the white and gray areas) in both nuclei.  The transition
densities and currents for the RPA modes with the maximum $B$(E3) at the
touching point are depicted in Fig.~\ref{den} (b) and (d).  These
figures indicate that the octupole collectivities of both $^{16}$O and
$^{208}$Pb are considerably diminished by each colliding partner.

\begin{figure}[htbp]
 \includegraphics[keepaspectratio,width=\linewidth]{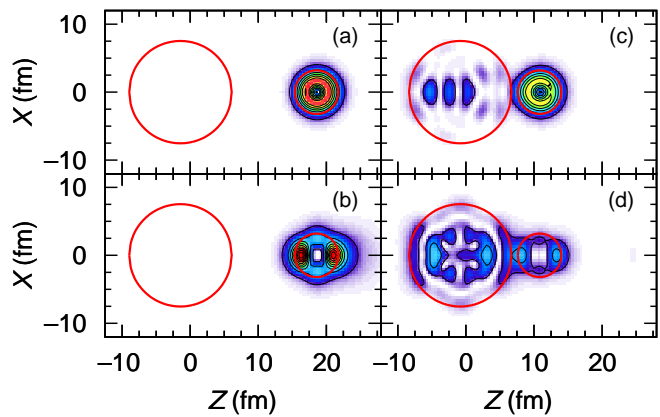}
 \caption{(Color online) Density distributions of the $p_{1/2}$ and
 $d_{5/2}$ states originally belonging to $^{16}$O at $R=20$ fm and
 their evolutions at $R=11.65$ fm.  The (red) thick solid circles
 indicate the half depth of the mean-field potential.  The contour lines
 correspond to multiples of 0.0015 fm$^{-3}$.  The colors are normalized
 at 0.01 fm$^{-3}$.}  \label{wf}
\end{figure}
 
The microscopic origin of the damping of these vibrations is easily seen
as follows.  At $R=20$ fm, the main proton components of the $3^-$ modes
are the excitations $p_{1/2}$ $\rightarrow$ $d_{5/2}$ and $p_{3/2}$
$\rightarrow$ $d_{5/2}$ for $^{16}$O, and the excitations $d_{3/2}$
$\rightarrow$ $h_{9/2}$ and $s_{1/2}$ $\rightarrow$ $f_{7/2}$ for
$^{208}$Pb [see the (red and blue) arrows around $R=19$ fm in
Fig.~\ref{nil}].  The density distributions of the $p_{1/2}$ and
$d_{5/2}$ states in $^{16}$O are displayed in (a) and (b) of
Fig.~\ref{wf}.  Their wave functions suffer major modifications near the
touching point at $R=11.65$ fm, as depicted in (c) and (d) of
Fig.~\ref{wf} [see also the (red) arrow at $R=11.65$ fm in
Fig.~\ref{nil}].  We can clearly see the neck formations in (d).  Also
for $^{208}$Pb, similar drastic changes of single-particle wave
functions occur for both protons and neutrons near the Fermi surface,
causing the damping of the collectivity of the $3^-$ vibration [see the
(blue) arrows around 10.6 fm in Fig.~\ref{nil}].

Finally, to see the correlation with the damping factor
phenomenologically introduced in the CC calculation, we compare the
calculated results with the damping factor that well reproduced the
experimental data of the fusion cross section for $^{208}$Pb + $^{16}$O
\cite{ich09}.  The damping factor is given by
$\Phi(r,\lambda_\alpha)=e^{-(r-R_d-\lambda_\alpha)^2/2a_d^2}$ for
$r<R_d+\lambda_\alpha$ (otherwise $\Phi=1$), where $a_d$ and $\lambda_a$
denote the damping width and the eigenvalues of the coupling matrix
elements, respectively.  The parameter $R_d$ is given by
$R_d=r_d(A_T^{1/3}+A_P^{1/3})$, where $r_d$ denotes the damping radius
parameter, and $A_T$ and $A_P$ the mass numbers of the target and the
projectile, respectively. In the calculation of Ref.~\cite{ich09},
$r_d=1.298$ fm and $a_d=1.05$ fm are used. Then, the largest eigenvalue
of $\lambda_\alpha$ is 1.46 fm.  In Figs.~\ref{damping} (a) and (b), the
dashed curves represent the damping factor with these parameters
normalized at $R=20$ fm.  We can see that the damping factor strongly
correlates with the calculated energy-weighted sums of $B$(E3) in the
low-energy region, which clearly indicates that the damping of the
quantum vibrations indeed occurs when the colliding nuclei approach each
other.

In summary, we have demonstrated the damping of the quantum octupole
vibrations of both $^{16}$O and $^{208}$Pb, when they approach each
other. To show this, we, for the first time, applied the RPA method to
the heavy mass-asymmetric $^{208}$Pb + $^{16}$O system.  We have
discussed the Nilsson diagram as a function of the center-of-mass
distance $R$ and have shown that the single-particle energies in
$^{16}$O are largely sifted to the positive-energy direction by the
strong Coulomb effects from the heavy-mass $^{208}$Pb in a colliding
process.  We calculated the $B$(E3) strengths for $^{16}$O and
$^{208}$Pb as a function of $R$.  The obtained $B$(E3) strengths are
substantially damped near the touching point of the colliding nuclei.
The obtained energy-weighted sum of $B$(E3) in the low-energy region
exhibits a strong correlation with the damping factor that reproduces
well the experimental data of the fusion cross section for $^{208}$Pb +
$^{16}$O. This is a clear evidence that the damping of the quantum
octupole vibrations indeed occur near the touching point in the deep
sub-barrier fusion reactions.  The drastic change of single-particle
wave functions consitituting the low-energy collective excitations
discussed in this paper would commonly occur in all deep sub-barrier
reactions. Therefore, the damping of quantum vibrations in both the
target and the projectile near the touching point seems to be a
universal mechanism causing the fusion hindrance, which should be taken
into account in the standard CC model.

A part of this research was funded by the MEXT HPCI STRATEGIC
PROGRAM. This work was supported by JPSJ KAKENHI Grant Number 15K05078.

\end{document}